# A Computational Model of Force within the Ligaments and Tendons in Progressive Collapsing Foot Deformity

Hamed Malakoutikhah, Erdogan Madenci, L. Daniel Latt
University of Arizona, Tucson, AZ

**ABSTRACT**

**Background:** Progressive collapsing foot deformity results from degeneration of the ligaments and the posterior tibial tendon (PTT). Our understanding of the relationship between the failures of them remains incomplete. We sought to improve this understanding through computational modeling of force in these soft tissues.

**Methods:** The impact of PTT and ligament tears on force changes in the remaining ligaments was investigated by quantifying ligament force changes during simulated ligament and tendon cutting in a previously validated finite element model of the foot. The ability of the PTT to restore foot alignment in a collapsed foot was evaluated by increasing the PTT force in a foot with attenuated ligaments and comparing the alignment angles to the intact foot.

**Results:** Rupture of any one of the ligaments led to overloading the remaining ligaments, except for the plantar naviculocuneiform, first plantar tarsometatarsal, and spring ligaments, where removing one led to unloading the other two. The attenuation of the plantar fascia, long plantar, short plantar, and spring ligaments significantly overloaded the deltoid and talocalcaneal ligaments. Isolated PTT rupture had no effect on foot alignment, but did increase the force in the deltoid and spring ligaments. Increasing the force within the PTT to 30% of body weight was effective at restoring foot alignment during quiet stance, primarily through reducing hindfoot valgus and forefoot abduction as opposed to improving arch height.

**Conclusion:** The attenuation of any one ligament often leads to overload of the remaining ligaments which may lead to progressive degeneration. The PTT can maintain alignment in the collapsing foot, but at an increased load which could lead to its injury. Early intervention, in the form of ligament repair or reconstruction, might be used to prevent the progression of deformity. Moreover, strengthening the PTT through therapeutic exercise might improve its ability to restore foot alignment.

## 1. INTRODUCTION

Progressive collapsing foot deformity (PCFD), initially known as adult acquired flatfoot deformity (AAFD), is a common degenerative disorder that encompasses a wide range of deformities,[37] most frequently longitudinal arch height loss, hindfoot valgus, and forefoot abduction. It results from the degeneration of both ligaments and tendons that support the arch and stabilize the foot.[29,37] Ligaments act as passive stabilizers and provide most of the support during quiet stance;[3,34,37] whereas the posterior tibial tendon (PTT) provides active foot stability during walking.[3,43,48] Active stability is created as the PTT inverts the midfoot on the hindfoot during midstance, effectively locking the transverse tarsal joint and creating a rigid medial column lever which can transfer force to the ground during terminal stance.[7,43,50]

Failure of any of the soft tissue stabilizers might change the force pattern within the other stabilizers, which would make them more prone to degeneration and failure over time.[23,43,45,53,57] An understanding of the relationship between the failure of the certain soft tissue stabilizers and changes in the tensile force within the remaining stabilizers would aid our understanding of patterns of degeneration and deformity. Such understanding could be used to guide the choice of reconstructive procedures and might help to develop strategies for early intervention that could be used to prevent the progression of deformity.

Although several experimental studies have been conducted on PCFD,[2,12,26,30] changes in the pattern of forces within the soft tissues have not been characterized. Cadaveric studies on forces in the soft tissues have mostly been limited to the plantar fascia, long plantar ligament, ankle ligaments, and Achilles tendon, whose length and accessibility allow them to be tested using direct and indirect methods of force measurements.[10,15,24,40,59] The difficulty in measuring forces and deformations within the foot from experimental models,[9,62] has led to a rise in the use of computational models such as finite element models to study PCFD.[8,27,34,55,60] However, only two of the studies in PCFD focused on forces and stresses in the soft tissues.[44,52] Unfortunately, the models used in both of these studies contain simplifications that limit their ability to make physiologically realistic predictions. Specifically, these models were only limited to four ligaments (the plantar fascia, long and short plantar, and spring ligaments), used linear materials, and had unrealistic boundary conditions due to the lack of a tibia and encapsulating soft tissue (EST).

The effect of the PTT on the stability of the foot has also been investigated using biomechanical cadaveric studies.[14,28,30] However, the results of these studies are contradictory. In addition, none of them have determined how much force in the PTT is required to maintain alignment in the collapsing foot or which components of the deformity are reduced by increasing force within the PTT. A computational model of the foot provides an ideal laboratory for gradually increasing the force in the PTT while monitoring changes in the foot configuration to evaluate the ability of the PTT to restore foot alignment for all components of the deformity and different scenarios of torn and attenuated ligaments.

More recently, a finite element model of the foot and ankle was developed and validated for both the neutrally aligned and collapsed foot to investigate the immediate effects of ligament tears on each component of the deformity.[34] In this study, it was found that the failure of any single ligament was not sufficient to cause immediate deformity. We hypothesized that although the failure of a single ligament does not cause deformity, it may lead to changes in the tension in the remaining ligaments, which could lead to their degeneration and the development of deformity over time. In addition, we hypothesized that the PTT may become injured secondarily as it attempts to maintain stability in a foot with attenuated ligaments. The current study uses our previously validated model to: (1) evaluate the impact of ligament attenuation or PTT insufficiency on the tensile force within the remaining soft tissues, and (2) determine the ability of the PTT to maintain foot stability when the principle ligamentous stabilizers are attenuated.

## 2. MATERIALS AND METHODS
### 2.1. Model reconstruction and mesh creation
The computational model was reconstructed using CT scan images of the left foot of a 50-year-old female cadaver with a weight of 600 N and a height of 160 cm who had no reported deformity or arthrosis[34] (Figure 1a, b). Briefly, pixel intensity threshold ranges were manually set to 648 mg/mL[56] in MIMICS V.12.1 (Materialize Inc., Leuven, Belgium) to separate the cortical from the trabecular bone. After simplifying the surface of the bones and EST in terms of the number of facets (Figure 1c) in MeshLab V.2016 (ISTI-CNR, Rome, Italy), the model was imported into ANSYS SpaceClaim V.2019 R1 (ANSYS Inc., Canonsburg, Pennsylvania, United States), where the irregular surface mesh was enhanced by reducing the element aspect ratio (Figure 1d). The cartilage was created by extruding specific surfaces of the bones in the normal direction. The location of the bones within the EST was determined using the average thickness of the first metatarsal fat pad and the heel fat pad, which were 6.75 mm and 19.94 mm, respectively.[4] Tension-only spring elements were used to model the ligaments in ANSYS. The finite element model of the foot was completed with 28 bones, 72 ligaments, cartilage, and the EST.[34]

The reconstructed geometry was meshed with 608,773 tetrahedral elements in ANSYS. Element sizes of 2 mm, 1 mm, 3 mm for the bones, cartilage, and EST were obtained, respectively, using a mesh convergence study conducted on the outcome parameters with a 10% reduction in element size.[58]

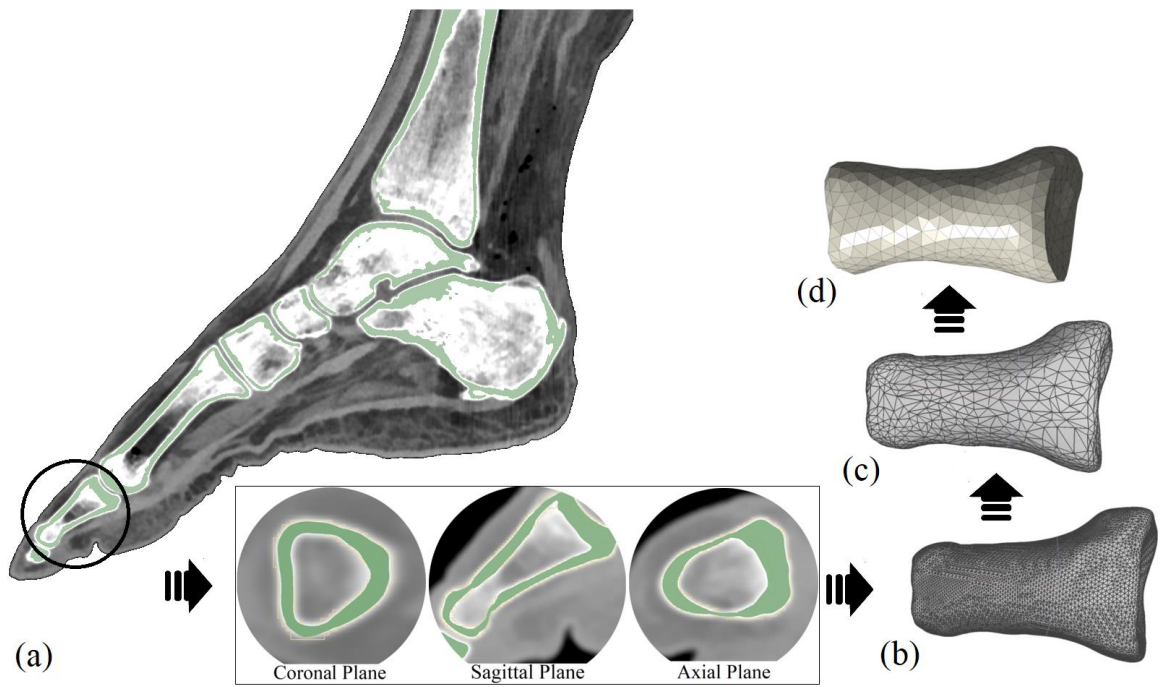

**Figure 1.** Reconstruction of a bone from CT scan images, a) separating the cortical from the trabecular bone by setting pixel intensity threshold, b) reconstructing the cortical bone, c) simplifying the bone surface in terms of the number of facets, d) creating mesh with aspect ratio close to one.

### 2.2. Mechanical properties of materials

The EST and cartilage were modeled using the hyperelastic Mooney-Rivlin material model because of their anisotropic properties and large nonlinear deformations.[1,33,34] Homogeneous and linear isotropic material properties were selected for the cortical and trabecular bones, and the ground.[34,49,52] The stiffness of the ligaments ($k = AE/L$), which accounts for both the material properties (elastic modulus E) and the geometry (cross-sectional area A and length L), was determined using published data.[32,34,51]

### 2.3. Load conditions

To simulate quiet stance, half the body weight was applied as the ground reaction force at the center of gravity (60 mm from the lateral malleolus and 67 mm from the fifth metatarsophalangeal joint).[39] A quarter of the body weight was applied at the Achilles insertion. The remaining tendon forces were determined from previously published data[36,47] and applied along tendon lines of action after being adjusted for a body weight of 600 N (Table 1). To simulate midstance, all forces were doubled.[36] Bonded (glued) contacts were used between the cortical and trabecular bones, and between the bones and cartilage. Nonlinear frictional contacts with coefficients of friction of 0.6 and 0.01 were used between the EST and the ground, and between the cartilage, respectively.[17,64] The proximal tibia was fixed only in the z-direction (vertical), allowing it to translate and rotate in all other degrees of freedom. This allowed for hindfoot valgus when simulating PCFD (Figure 2).

**Table 1.** Tendon forces during quiet stance calculated based on the body weight of 600 N.[35]

| Tendon | Force (N) | Tendon | Force (N) |
| --- | --- | --- | --- |
| Posterior tibial (PT) | 25.71 | Flexor hallucis longus (FHL) | 12.86 |
| Anterior tibial (AT) | 0 | Flexor digitorum longus (FDL) | 6.43 |
| Peroneus longus (PL) | 20.57 | Extensor hallucis longus (EHL) | 12.86 |
| Peroneus brevis (PB) | 10.28 | Extensor digitorum longus (EDL) | 6.43 |

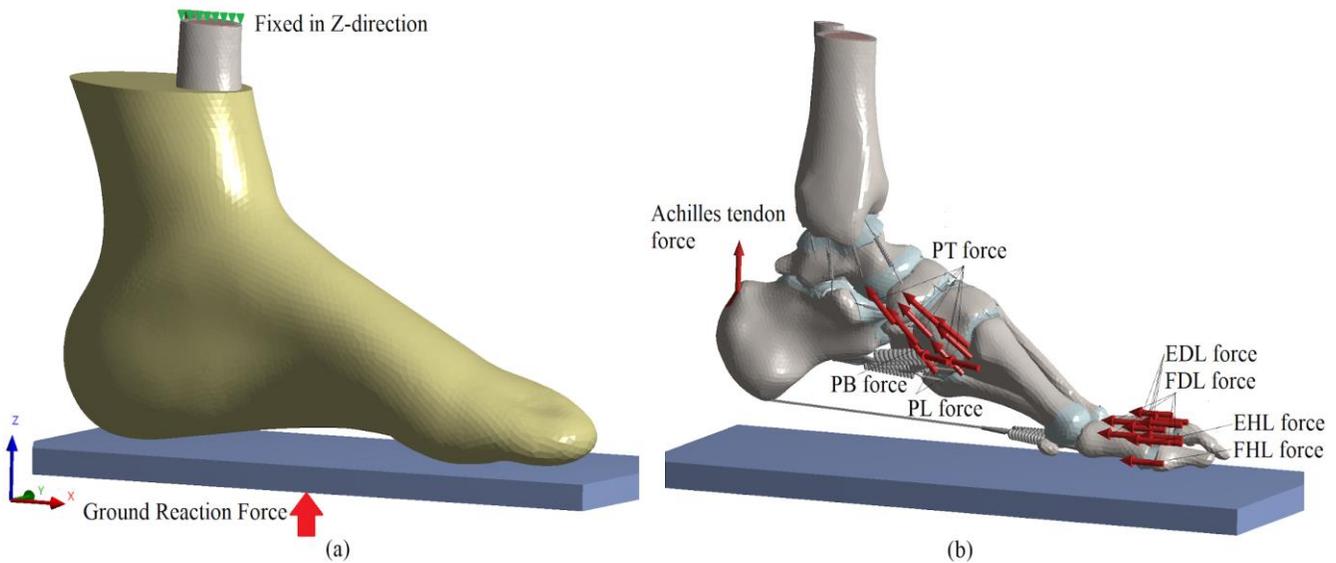

**Figure 2.** a) Ground reaction force and boundary conditions, b) tendon forces.

## 2.4. Ligament failure and collapsed foot simulation

The ligaments involved in PCFD were identified through a number of experimental and computational studies[2,12,34] (Table 2). The degeneration of these ligaments and tibialis posterior tendon dysfunction (PTTD) are the primary causes of collapsing foot deformity.[29,34,37] Specifically, ligaments attenuate over time as a result of repeated loading on the arch, causing them to become elongated and thus nonfunctional.[18,41] To simulate torn or elongated ligaments, the ligaments must be either removed or pre-stretched to specific lengths.[34] Thus, we removed all of the involved ligaments (Table 2) and unloaded the PTT to simulate the collapsed foot.

**Table 2.** Ligaments of interest in PCFD

| Ligament | Ligament |
|---|---|
| Plantar fascia (PF) | Deltoid ligament (DL) |
| Long plantar ligament (LPL) | ➢ Deep layer (DDL) |
| Short plantar ligament (SPL) | ○ Posterior tibiotalar ligament (PTTL) |
| Spring ligament (SL) | ○ Anterior tibiotalar ligament (ATTL) |
| Interosseous talocalcaneal ligament (ITCL) | ➢ Superficial layer (SDL) |
| Cervical ligament (CL) | ○ Tibiospring ligament (TSL) |
| (plantar) Naviculocuneiform ligament (NCL) | ○ Tibionavicular ligament (TNL) |
| (1st plantar) Tarsometatarsal ligament (TMTL) | ○ Tibiocalcaneal ligament (TCL) |
| Plantar cuneocuboid ligament (PCCL) | ○ Superficial posterior tibiotalar ligament (STTL) |

## 2.5. Foot model validation and sensitivity analysis

Both normal and collapsed foot models were validated using clinically relevant kinetic and kinematic measures reported for normal foot and flatfoot populations including (1) von Mises stress in the metatarsal bones, (2) plantar pressure distribution, (3) PF force, (4) joint contact characteristics, (5) four widely used foot alignment angles, and (6) five anatomical distances.[34] The validation was considered successful if the predicted and reported outcome measures differed by less than 10%.[63] The predicted PF force was compared to the value obtained from the following formula, which was derived from a linear regression analysis used to calculate the PF force as a function of bodyweight and Achilles tendon force.[15]

$$\text{PF force} = 0.474 \times \text{Achilles tendon force} + 0.041 \times \text{Body weight}$$
$$= 0.474 \times 150 \text{ (N)} + 0.041 \times 600 \text{ (N)} = 95.7 \text{ (N)} \qquad (1)$$

Due to differences in ligament stiffness between reported data,[27,49] the sensitivity of outcome measures (force in the ligaments and foot alignment angles) to changes in ligament stiffness was investigated by analyzing outcomes for ligaments with stiffness reduced by half and ligaments with stiffness increased to infinity. The foot model was also subjected to a sensitivity analysis with respect to uncertain insertion points of the ligaments which involved translating them by 3 mm in random directions.

## 2.6. Evaluation of the impact of ligament/ tendon attenuation on force within the remaining ligaments

The impact of individual ligament or a specific combination of ligaments tears on the force in the remaining ligaments was investigated by comparing the force values of each case to an intact foot. Individual ligament attenuation was simulated by removing one ligament at a time while the others remained intact. Specifically, we evaluated the attenuation of the plantar ligaments (PF, SPL, and LPL) in isolation and in combination with DL and SL because of their critical role in maintaining foot alignment.[34] The effect of the PTT tear on foot alignment and force changes within the ligaments was then assessed by unloading the PTT while the ligaments were intact.

The Meary's angle (MA) and calcaneal pitch angle (CPA) were used to quantify arch collapse (Figure 3a). To evaluate hindfoot valgus (Figure 3b) and forefoot abduction (Figure 3c), the hindfoot alignment angle (HAA) and talonavicular coverage angle (TCA) were used, respectively.

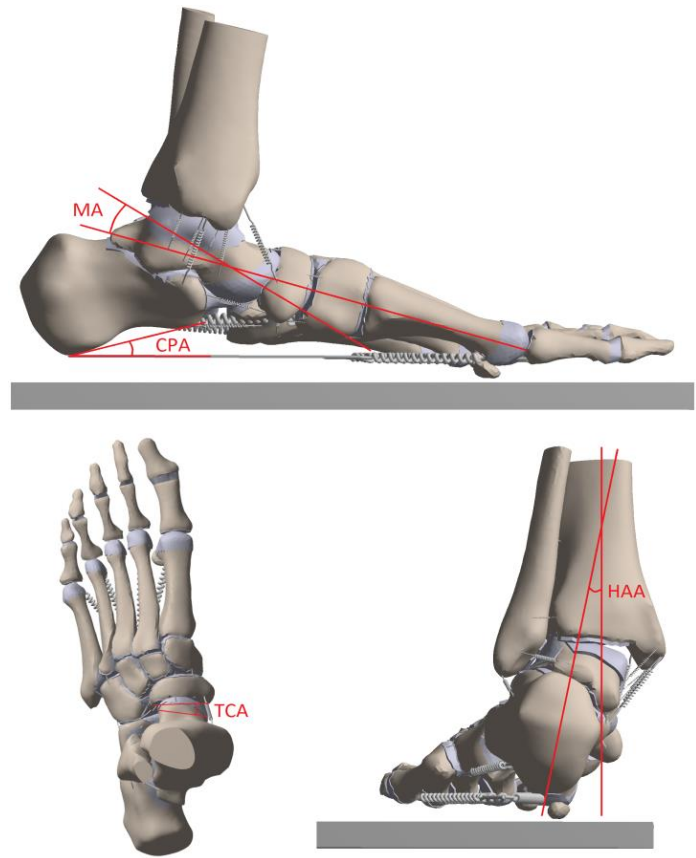

**Figure 3.** Foot alignment angles: a) medial view, b) posterior view, c) AP view. *MA*: Meary's angle (the sagittal plane angle between the long axis of the talus and the first metatarsal), *CPA*: calcaneal pitch angle (the sagittal plane angle between the line along the inferior inclination axis of the calcaneus and the horizontal plane), *HAA*: hindfoot alignment angle (the coronal plane angle between the axis of the calcaneal tuberosity and the longitudinal axis of the tibia), and *TCA*: talonavicular coverage angle (the transverse plane angle formed by the lines of articular surfaces of the navicular and talus).

## 2.7. Evaluation of the PTT capability to restore foot alignment

The failure of foot-locking is one of the most common problems in PCFD during gait.[54] As the primary dynamic stabilizer of the foot, the PTT may become overloaded to compensate for the loss of the ligamentous stabilizers by partially locking the transverse tarsal joint and creating a rigid medial column lever. To evaluate the ability of the PTT to restore foot alignment when the primary ligamentous stabilizers are attenuated, we gradually increased the force within the PTT in a foot with attenuated ligaments and compared the measured foot alignment angles to the intact foot.

## 3. RESULTS

### 3.1. Impact of isolated and combined ligament failures on the remaining ligaments

In the sensitivity analysis of the outcome measures to ligament stiffness and insertion points, we found less than 6% changes in the force within the ligaments and foot alignment angles. In the investigation of the impact of isolated ligament failures on force within the remaining ligaments (Figure 4), we found that the loss of the PF can significantly increase the force in the remaining ligaments except the DDL. The failure of the LPL mostly raised the force in the PF, SL, and SPL. Similarly, as the SPL failed, the PF, SL, and LPL experienced a rise in force. Removing the SL incremented the force in the PF, LPL, SPL, and ITCL but unexpectedly decreased the force in the NCL and TMTL. The failure of either the NCL or TMTL increased the force in the PF while it reduced the force in the SL. However, the isolated rupture of the DL, ITCL, or CL had no effect on the remaining ligaments.

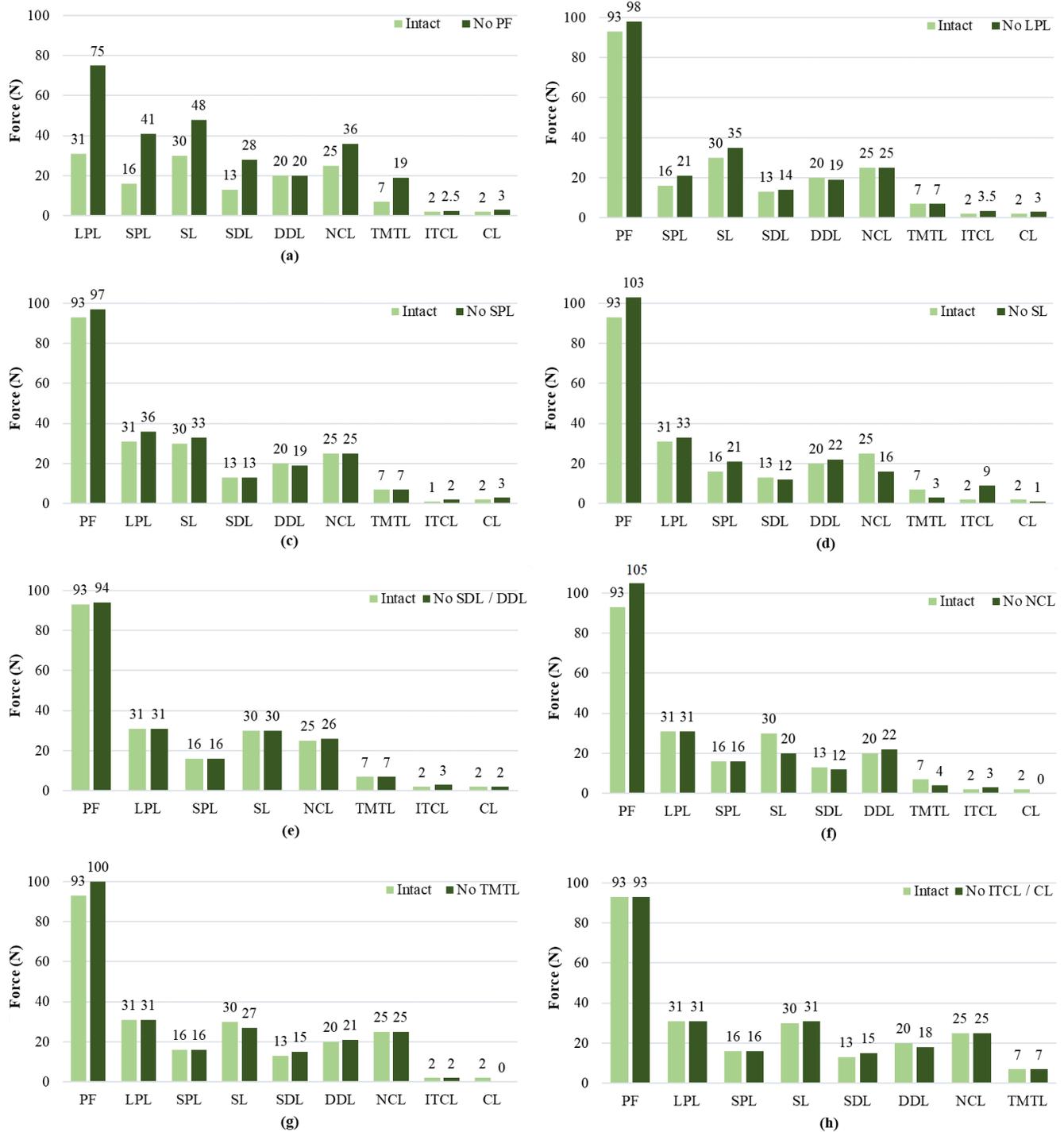

**Figure 4.** Comparison of the force within the ligaments between the model of the intact foot and the foot with simulated rupture of a) PF, b) LPL, c) SPL, d) SL, e) DL, f) NCL, g) TMTL, h) ITCL and CL.

The study of different combinations of the primary ligamentous arch stabilizers, the plantar ligaments (PF, LPL, and SPL), SL, and DL, showed that the loss of the plantar ligaments could increase the force in all the remaining ligaments except the DDL (Figure 5a). When compared to the foot with only plantar ligament rupture, removing the SL along with the plantar ligaments resulted in a significant increase in the force in the SDL, DDL, and ITCL but a decrease in the force in the NCL and TMTL (Figure 5b). However, failure of the DL along with the plantar ligaments increased the force in all the remaining ligaments (Figure 5c). Finally, the failure of the SL and DL, as well as the plantar ligaments, led to a considerable increase in the force in the ITCL (Figure 5d). The loss of other ligaments such as the NCL, TMTL, ITCL, or CL along with the plantar ligaments had a minor impact on the force in the remaining ligaments.

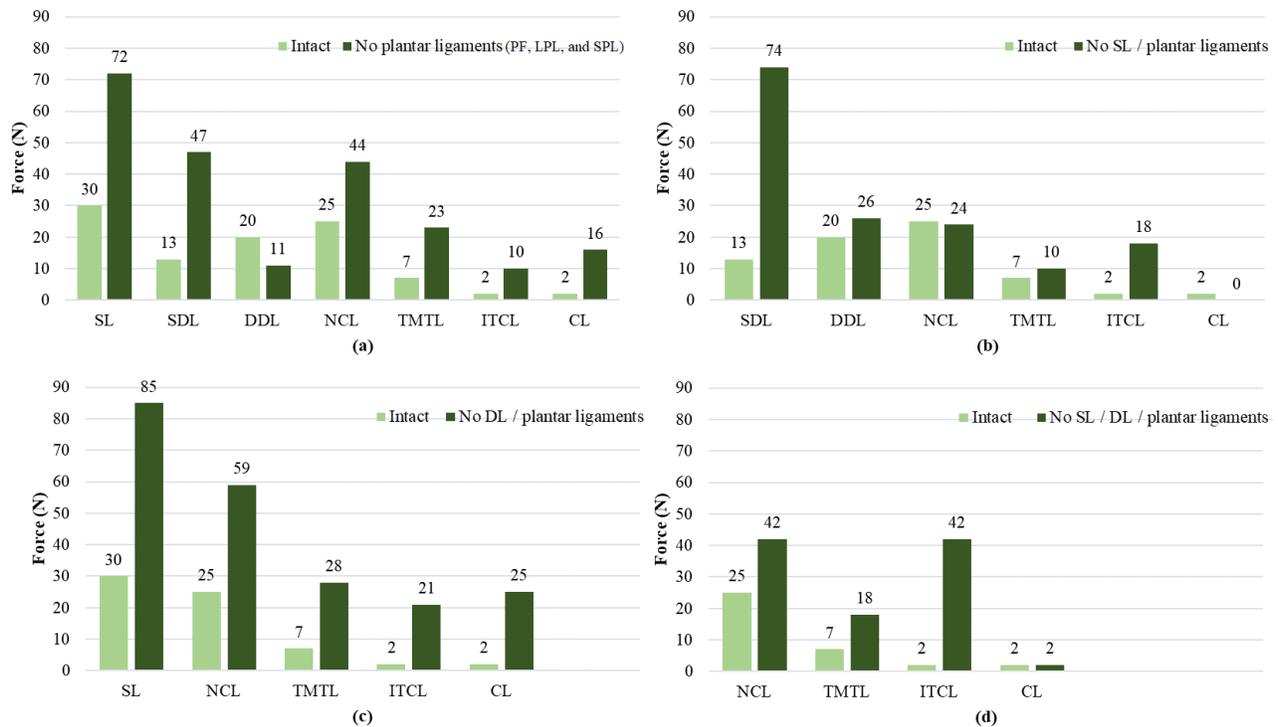

**Figure 5.** Comparison of the force in the ligaments between the model of the intact foot and the foot with simulated rupture of a) the plantar ligaments (PF, LPL, and SPL), b) SL along with the plantar ligaments, c) DL along with the plantar ligaments, d) SL and DL along with the plantar ligaments.

### 3.2. Impact of PTT rupture on the force within the ligaments

Isolated PTT rupture had no effect on foot alignment as the MA, CPA, HAA, and TCA values of 1.9, 22, -0.35, and 0.25 degrees were unchanged from the intact foot. Isolated rupture of the PTT increased the force in the SL and DL while the force in the plantar ligaments remained relatively unchanged compared to the intact foot (Figure 6).

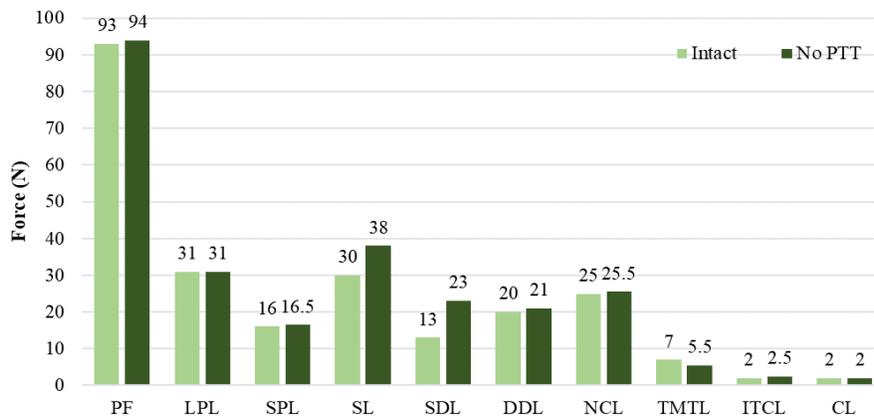

**Figure 6.** Comparison of the force in the ligaments between the foot with ruptured PTT and the intact foot.

### 3.3. Ability of the PTT to restore alignment in the foot with attenuated ligaments

We found that increasing the PTT force to 180 N, which is 30% of body weight (180 N/ 600 N = 0.3), led to a nearly complete restoration of foot alignment in the collapsed foot during quiet stance (Figure 7). This is nearly seven times the normal force within the PTT. The angles associated with arch height (MA and CPA) were restored to a lesser extent than those associated with hindfoot and forefoot alignment (HAA and TCA), indicating that the PTT may be more able to restore alignment in certain types of PCFD than others. In the foot with simulated rupture of the DL and plantar ligaments, increasing force within the PTT by 20% of body weight led to a restoration of near normal hindfoot alignment (Figure 7c). Whereas in the foot with simulated rupture of the SL and plantar ligaments, the forefoot abduction deformity was completely compensated only when the PTT force was increased to 50% of body weight (Figure 7d). At midstance, these values were nearly doubled.

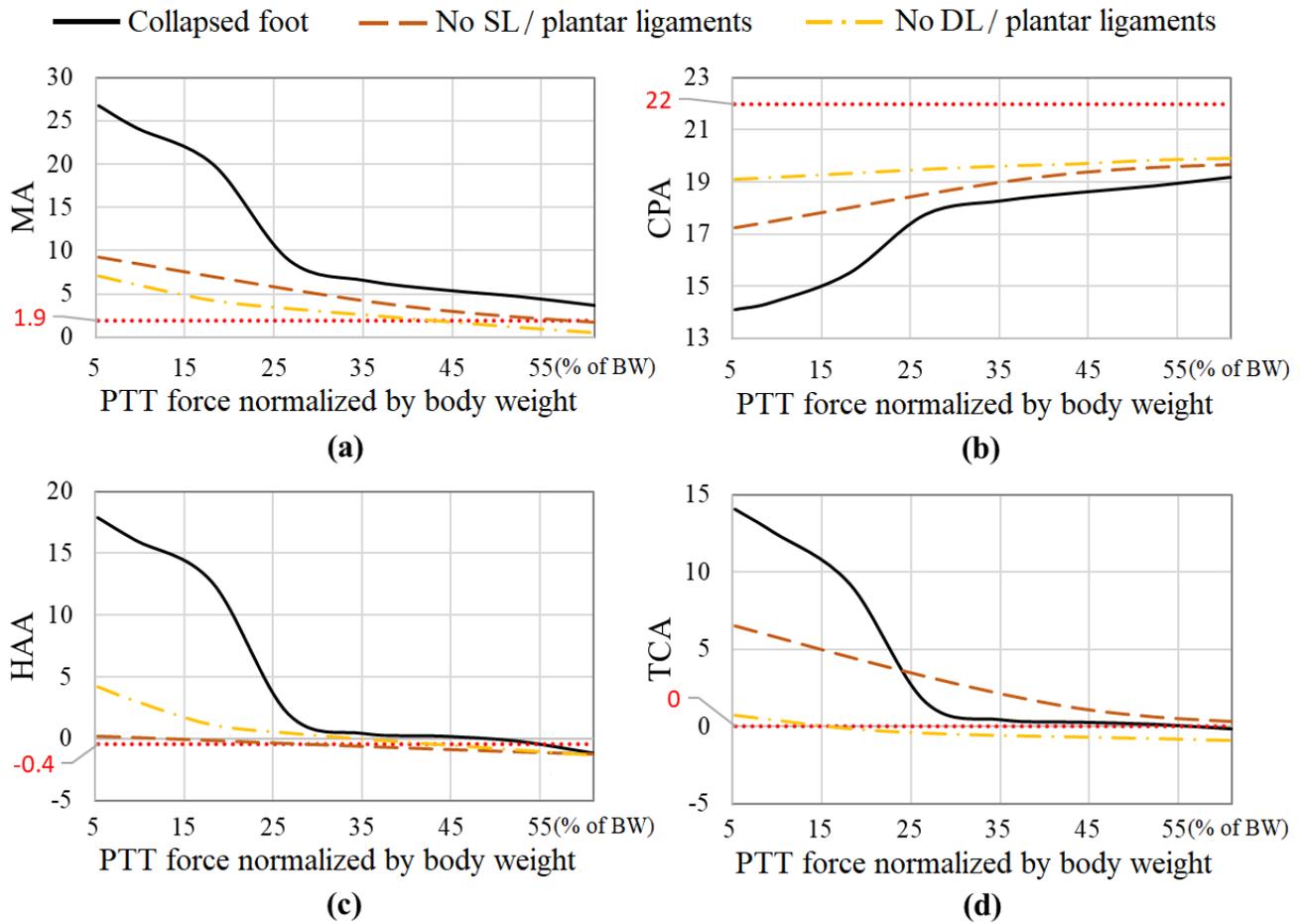

**Figure 7.** Restoration of foot alignment angles (degrees) during quiet stance with increasing force in the PTT for simulated rupture of all ligaments (solid line), SL and plantar ligaments (dashed line), DL and plantar ligaments (dashed dot line), compared to the intact foot (dotted line).

## DISCUSSION

Although the rupture of any individual ligament does not cause immediate deformity,[34] the remaining ligaments might be overloaded to compensate for the loss. Repetitive overload of the ligaments is known to result in fatigue, elongation, and attenuation that lead to the progression of deformity over time.[18,23,57] However, the strength of this effect may differ among ligaments depending on a particular ligament's anatomical position and corresponding role in maintaining foot alignment. In this study, we used a validated model of the neutrally aligned and collapsed foot to determine the impact of ligament rupture or PTT insufficiency on the tensile force within the remaining ligaments, and to evaluate the ability of the PTT to restore foot alignment in the setting of ligament attenuation.

Previous studies on the plantar ligaments (PF, LPL, and SPL) predicted that releasing one of them would result in an increase in force in the other two.[21,52] When we simulated isolated ligament ruptures, we found that the individual failure of any one of the plantar ligaments significantly increased the force in all of the remaining ligaments, including the other two, as well as the SL, DL, ITCL, CL, NCL, and TMTL. This was particularly true for the PF. We also found that rupture of the SL increased the force in the remaining ligaments, with the exception of the NCL and TMTL, which were partially unloaded. This surprising result suggests that if one of these three ligaments located in series along the medial column fails, the medial column becomes unloaded, reducing the force in the other two. Finally, isolated failure of the DL or talocalcaneal ligaments was found to have no impact on the force within the remaining ligaments.

When we simulated the rupture of combinations of the ligaments, we also found overloading of the remaining ligaments, with the exception of the SL, NCL, and TMTL (in which removing one led to unloading the other two). Moreover, we found that if the plantar ligaments (PF, LPL, and SPL) are intact, the DL will not become overloaded. In other words, a stable arch prevents hindfoot valgus. Alternatively, this could be expressed as "forefoot driven hindfoot varus", which plays a role not only in creating deformity in cavus foot, but in preventing deformity in flatfoot. In terms of PCFD, this means that class C deformity (medial column instability) may be a prerequisite for the development of class A deformity (hindfoot valgus). Additionally, failure of the DL combined with the plantar ligaments and the SL, created hindfoot valgus and arch collapse, which led to overload of the ITCL. In other words, attenuation of the plantar ligaments (PF, LPL, and SPL) and SL causes a sagging arch, which overloads the DL, leading to hindfoot valgus over time, which then overloads the ITCL and leads to subtalar joint malalignment.

When we simulated the isolated PTT rupture, we found no effect on foot alignment, indicating that a tear of the PTT cannot acutely lead to PCFD if the primary ligamentous stabilizers are intact. This is consistent with the clinical consensus that PTT rupture alone does not cause foot deformity.[37] However, rupture of the PTT was found to increase the force in the SL and DL, which may predispose them to attenuation and rupture. This is consistent with the experimental observations that SL tears[2,12,16,20,61] and DL tears[25] occur in patients with PTTD. In summary,

- An Isolated rupture of any one of the PF, LPL, or SPL increased the force in all the remaining ligaments.
- An Isolated rupture of any one of the SL, NCL, or TMTL increased the force in all of the remaining ligaments while decreasing the force in the other two.
- An isolated rupture of the DL or ITCL had no effect on the force in the remaining ligaments.
- An isolated rupture of the PTT increased the force only in the DL and SL.
- A combined rupture of the PF, LPL, SPL, and SL significantly increased the force in the DL, which may have resulted in hindfoot valgus over time, which subsequently increased the force in the ITCL.

Moreover, we found that almost 30% of body weight in the PTT (seven times the normal force within the PTT) was required during quiet stance to restore the foot with attenuated ligaments to neutral alignment. Augmentation of *in vivo* PTT force to this extent is unrealistic. However, we found that a lesser increase in PTT force provides partial compensation of the alignment. Because the PTT can partially restore foot alignment, it may become secondarily injured when the ligaments are attenuated or torn. Moreover, strengthening of the PTT through therapeutic exercise can improve its ability to restore foot alignment and should be a cornerstone of non-operative treatment of PCFD. These findings support those of cadaveric studies, which found that overloading the PTT by 150% was unable to fully compensate for isolated SL insufficiency despite visible position changes toward restoring the arch alignment.[28,30] On the other hand, a biomechanical cadaveric study[14] suggested that actuation of PTT had no effect on restoring foot alignment. However, it is likely that the foot alignment angles remained unchanged in their study due to some limitations, such as the absence of other tendons, particularly the Achilles tendon, the use of normal foot cadavers, and the fixation of the proximal tibia, all of which prevented the development of hindfoot valgus and arch collapse in their initial model. Furthermore, we found that increasing PTT force primarily reduced hindfoot valgus and forefoot abduction while having a smaller impact on arch height.

This likely occurred because the PTT courses behind the medial malleolus and inserts on the plantar aspects of the navicular and medial cuneiform, creating an effective moment arm about the talonavicular joint in the transverse and coronal planes, but not in the sagittal plane. This agrees with the clinical observations that the PTT primarily assists the ligaments in producing hindfoot inversion[2] and that patients with PTT and SL injuries have severe hindfoot abnormalities.[20]

Although it has been classically thought that PCFD or AAFD occurs secondary to PTTD,[6,19,20,22,31,35,38] it has also been hypothesized that PCFD triggered by the attenuation or rupture of the SL or DL leads to increased strain on the PTT and the progression of PTTD.[11,13,42] The current work provides support for both of these potential mechanisms because when one of the SL, DL, or PTT fails, the remaining ones will be overloaded, leading to degeneration and eventual failure over time.[23,43,45,57] In other words, PTTD could be either primary or secondary to collapsing foot deformity caused by ligament attenuation. Our findings indicate that early detection and intervention might be used to prevent the development of deformity and symptomatic disease.

The model used in this study differs in several ways from previously published models[44,52] that have explored biomechanical parameters in flatfoot. First, the current model included all the named ligaments of the foot and ankle, whereas previous models included only four primary ligaments. Second, unlike previous models, we used nonlinear and anisotropic materials that closely mimic the hyperelastic properties of real biologic materials. Moreover, previous models oversimplified the boundary conditions (e.g. fixing the calcaneus and metatarsal heads to the ground support or fixing the tibia in all degrees of freedom). These oversimplifications led the foot models to be unable to adequately deform to simulate the changes in alignment angles that occur with flatfoot. For example, Portilla et al.[44] reported an internal Moreau-Costa-Bertani angle of 130 and a Kite angle of 23 degrees for flatfoot, both of which are out of the reported clinical ranges for these angles, which are $144 \pm 9$ and >40 degrees, respectively.[5,16] A third difference pertains to the elements chosen to model ligaments and tendons. We modeled ligaments as tension-only spring elements which only support tensile forces because this reflects the mechanical behavior of real ligaments and tendons. In contrast, Portilla's model used solid elements, which are capable of resisting bending moments and compressive forces. This unrealistic choice of element types creates very high non-physiologic stresses, especially in curved structures such as the PTT, when they are subjected to a tensile load. Finally, the present model was validated for both normal and flatfoot conditions, which had not been done in either of the previous models.

Our model has several limitations. First, the model was created from CT scan images of a cadaveric foot in a supine position. The insertions of the ligaments and location of the bones during standing were obtained from the literature. In contrast, an MRI based solid model would have provided more information about the soft tissues. Alternatively, a weightbearing CT could have provided information about the position of the bones during stance. Moreover, in this study, the ligaments were modeled as linear elastic springs. However, this simplification may only result in negligible errors since ligaments experience very small deformation during working conditions and materials deform linearly with applied force under small deformations.[46]

**CONCLUSION**

A validated finite element model of the neutrally aligned and collapsed foot was used to evaluate the impact of ligament attenuation or PTT insufficiency on the tensile force within the remaining soft tissue stabilizers, and to determine the ability of the PTT to maintain foot alignment when the principle ligamentous stabilizers are attenuated. We found that depending on which ligaments had failed, isolated or combined ligament ruptures resulted in varying changes in force in the remaining ligaments. We also found that in the presence of ligament attenuation, nearly 30% of body weight (seven times the normal force within the PTT) was required to maintain foot alignment during quiet stance, whereas a smaller increase in PTT force provided partial compensation of the alignment, primarily by reducing hindfoot valgus and forefoot abduction. The results of this study might be used to help develop novel strategies for operative treatment of flatfoot using ligament reconstructions as opposed to osteotomies, and also emphasize the importance of strengthening the PTT through therapeutic exercise to improve its ability to restore foot alignment.

## ACKNOLEDGEMENTS
We would like to thank Paragon 28, Inc. for financial support.

## CONFLICT OF INTERESTS
This work was partially supported by a research grant from Paragon 28, Inc.